\documentclass[runningheads]{llncs}
\usepackage[T1]{fontenc}
\usepackage{graphicx}
\usepackage{booktabs}
\usepackage{caption}
\usepackage{adjustbox}

\usepackage{booktabs}
\usepackage{caption}
\usepackage{adjustbox}
\usepackage[most]{tcolorbox}
\usepackage{listings}
\usepackage{xcolor}
\lstdefinelanguage{gherkin}{
  morekeywords={Scenario, Given, When, Then, And, But},
  sensitive=true,
}

\usepackage{listings}
\usepackage{xcolor}

\lstdefinelanguage{gherkin}{
  morekeywords={Scenario, Given, When, Then, And, But},
  sensitive=true
}

\lstdefinestyle{gherkin-style}{
  language=gherkin,
  basicstyle=\ttfamily\small,
  keywordstyle=\color{blue}\bfseries,   
  frame=single,
  breaklines=true,
  backgroundcolor=\color{gray!5},
  columns=fullflexible,
  keepspaces=true,
  captionpos=b,
  moredelim=[is][\color{gray}\itshape]{[}{]},   
  literate=
    {Vehicle.Cabin.ChildPresenceDetection.IsChildDetected}{{\color{red}Vehicle.Cabin.ChildPresenceDetection.IsChildDetected}}1
    {Vehicle.Cabin.Infotainment.HVAC.AutoOverrideActive}{{\color{red}Vehicle.Cabin.Infotainment.HVAC.AutoOverrideActive}}1
}

\begin{document}
\title{GenAI-based test case generation and execution in SDV platform}
%
%
\author{Denesa Zyberaj\inst{1}\orcidID{0009-0009-4207-8723}
\and
Lukasz Mazur\inst{2}\orcidID{0009-0002-0997-0997}
\and
Nenad Petrovic \inst{2}\orcidID{0000-0003-2264-7369}
 \and
Pankhuri Verma\inst{3}\orcidID{0009-0008-1741-4214}
 \and
Pascal Hirmer\inst{1}\orcidID{0000-0002-2656-0095}
 \and
Dirk Slama\inst{3}
\and
Xiangwei Cheng\inst{3}\orcidID{0000-0001-7219-3532}
\and
Alois Knoll\inst{2}\orcidID{0000-0003-4840-076X} 
}

\authorrunning{Zyberaj et al.}
%
\institute{
Mercedes-Benz AG, Germany \\
\email{denesa.zyberaj@mercedes-benz.com, pascal.hirmer@mercedes-benz.com}
\and
Technical University of Munich, Germany \\
\email{lukasz.mazur@tum.de, nenad.petrovic@tum.de, k@tum.de}
\and
Ferdinand-Steinbeis-Institut der Steinbeis-Stiftung, Germany \\
\email{dirk.slama@bosch.com, pankhuri.verma@ferdinand-steinbeis-institut.de, chris.cheng@ferdinand-steinbeis-institut.de
}
}

\maketitle              
\begin{abstract}
This paper introduces a GenAI-driven approach for automated test case generation, leveraging Large Language Models and Vision-Language Models to translate natural language requirements and system diagrams into structured Gherkin test cases. The methodology integrates Vehicle Signal Specification modeling to standardize vehicle signal definitions, improve compatibility across automotive subsystems, and streamline integration with third-party testing tools. Generated test cases are executed within the digital.auto playground, an open and vendor-neutral environment designed to facilitate rapid validation of software-defined vehicle functionalities. We evaluate our approach using the Child Presence Detection System use case, demonstrating substantial reductions in manual test specification effort and rapid execution of generated tests. Despite significant automation, the generation of test cases and test scripts still requires manual intervention due to current limitations in the GenAI pipeline and constraints of the digital.auto platform.


 \and digital.auto
\end{abstract}
\section{Introduction}
The transition to software-defined vehicles (SDVs) significantly increases software complexity, intensifying systematic testing challenges. Despite established standards like ASPICE, automotive test planning often remains inconsistent, manual, and poorly aligned with evolving requirements~\cite{Lami2016,Juhke2020,Ambrosio2017}.Requirements expressed in natural language further complicate automated test derivation~\cite{bruel2020,Altinger2014,Zhao2021}, limiting traceability~\cite{Garousi2017,Norheim2024}.
Recent advances in Generative AI (GenAI), particularly Large Language Models (LLMs) and Vision-Language Models (VLMs), offer potential solutions by translating ambiguous inputs into structured representations, critical for SDV development.

\begin{figure}[t]
    \centering
    \includegraphics[width=0.7\linewidth]{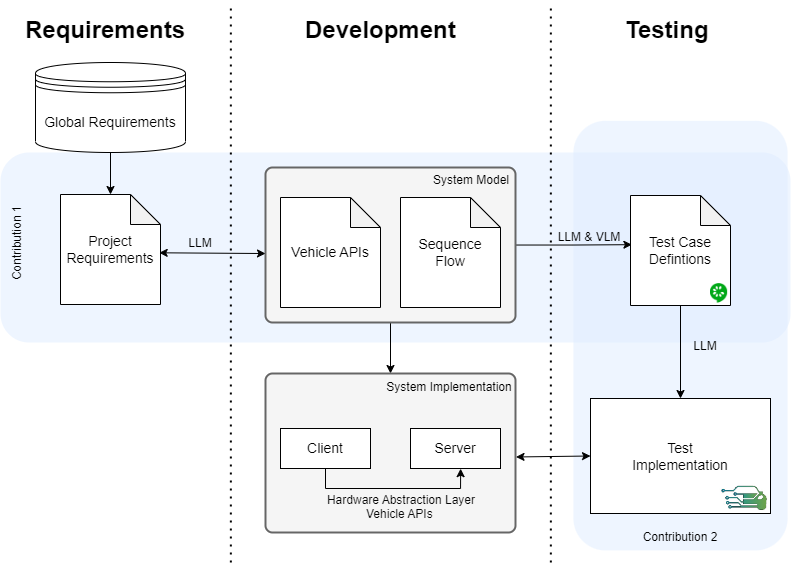}
    \caption{Overview of GenAI-driven test case generation leveraging LLMs, VLMs, and VSS integration for SDV testing automation}
    \label{fig:intro}
\end{figure}

In this work, as depicted in Fig.~\ref{fig:intro}, we propose a GenAI-based methodology leveraging LLMs and VLMs to automate the extraction and generation of structured test cases from automotive system requirements provided in natural language and diagrams into executable test cases written in Gherkin\footnote{Official Gherkin website: \url{https://cucumber.io/docs/gherkin/}} syntax. These are further translated into Python-based test scripts compatible with SDV testing environments .
Our methodology integrates Vehicle Signal Specification (VSS), a standardized data schema widely recognized across the automotive industry, to facilitate interoperability across diverse automotive subsystems and standardize the definition of vehicle signals. The entire pipeline is designed to integrate with the digital.auto \cite{slama2023digitalauto} playground, a vendor-neutral platform providing digital-twin environments that significantly accelerate the validation of SDV functionalities. We evaluate our approach using the Child Presence Detection System (CPDS), chosen due to its safety-critical nature and event-driven complexity, representing typical challenges faced in SDV testing.

The remainder of this paper is structured as follows: Section~\ref{method} details our proposed methodology and implementation. Results from our case study on CPDS are presented in Sect.~\ref{usecase}. We conclude with a discussion of our findings and future work in Sect.~\ref{conclusion}.

\section{Methodology and Implementation}
\label{method}
\subsection{Methodology}
Summarization, analysis, and text generation capabilities based on word sequence predictions by LLMs make them suitable for tasks such as VSS signal mapping based on a catalog of VSS signals available in digital.auto, as well as Python code generation based on test case specifications. On the other side, VLMs are leveraged in order to extract relevant information from UML sequence and state diagrams, building upon our work from \cite{Petrovic2025}. 

GenAI-driven test case generation consists of the following steps: 1) \textbf{Signal extraction} -- VLM interprets the diagram and extracts potential vehicle signals and working process used in communication and control, guided by the prompt; 2) \textbf{Mapping signals} -- LLM maps the identified signals from the sequence diagram to the existing VSS format (e.g., Branch, Sensor, Actuator); 3) \textbf{Gherkin test case generation} -- visual input (flowchart or state diagram) is used to generate Gherkin test cases relying on multimodal-enabled LLM solution (outside the scope of this paper); 4) \textbf{Python test file generation} -- using the refined VSS signals and test logic, the LLM generates an automated Python test file that simulates the test scenarios based on the sequence or state diagram. 
\begin{figure}[t]
    \centering
    \includegraphics[width=1\linewidth]{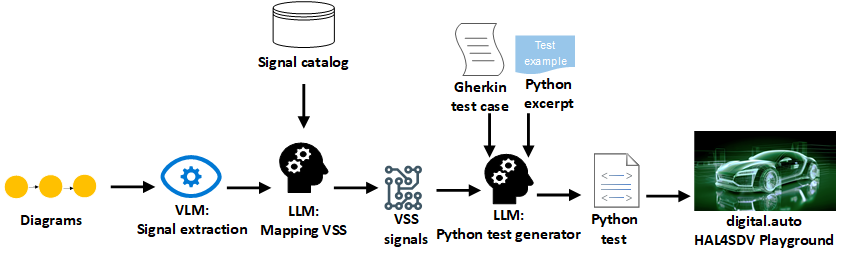}
    \caption{GenAI-driven test generation workflow leveraging VSS specification and Gherkin test cases.}
    \label{fig:genaiwf}
\end{figure}

\subsection{Implementation}
We utilized prompting strategies such as Chain of Thought and Few-Shot Learning. For VSS signal mapping, code generation, and Gherkin test case extraction, we used the LLM GPT-4o. 
The prompting process was structured into three distinct stages, each explicitly instructing the models to request clarification in case of ambiguities (see Table~\ref{tab:vss_tasks}). Initial prompts instructed models to generate clearly defined Given-When-Then test scenarios with explicit timing and conditional logic. Subsequent prompts required models to reference only valid VSS signals, explicitly requesting clarification when direct mappings were unavailable. Final-stage prompts directed models to strictly adhere to provided digital.auto playground templates, ensuring compatibility and readability through explicit comments and scenario referencing.

Initial outputs were manually validated at each stage to ensure accuracy and realism. This iterative refinement significantly improved reliability. Several technical challenges were identified throughout this process. Notably, ensuring the generated Python test scripts fully match the functional capabilities and constraints of the digital.auto platform, necessitating frequent manual adjustments.

\begin{table}[t]
\centering
\caption{Prompts overview}
\small
\begin{adjustbox}{max width=\textwidth}
\begin{tabular}{@{}p{4cm} p{10cm}@{}}
\toprule
\textbf{Task} & \textbf{Prompt} \\ 
\midrule
VSS Signal Extraction from Diagrams & 
\textit{You are an expert assistant trained in software architecture and automotive systems. Analyze the uploaded UML state machine diagram, which models the behavior of a component in an automotive software system. Please extract structured information in the format below. Make sure to identify states, transitions, composite (nested) states.} \\

Mapping VSS Signals & 
\textit{Based on list of identified signals [signal list from diagram] perform mapping with respect to list of possible signals [VSS catalog].} \\

Code generation for digital.auto & 
\textit{Generate Python test based on example [digital.auto test example] starting from Gherkin test case [Gherkin test case] taking into account list of available VSS signals [VSS signals].} \\
\bottomrule
\end{tabular}
\end{adjustbox}
\label{tab:vss_tasks}
\end{table}




\section{Use case: Child Presence Detection System}
\label{usecase}
A Child Presence Detection System (CPDS) detects and responds to unattended children in parked vehicles. We modeled a simplified CPDS based on Euro NCAP guidelines and relevant patents~\cite{euroncap,patent1,patent2}. 

CPDS employs direct sensors (cameras, UWB modules, microphones) and indirect sensors (seat pressure monitors, seatbelt indicators), while actuators include HVAC systems, horns, lights, and door locks. Due to this complexity, safety relevance, and its event-driven architecture, the CPDS is ideal for demonstrating structured, scenario-based test case generation using Gherkin~\cite{Zyberaj2025}.

\begin{figure}
    \centering
    \includegraphics[width=1\linewidth]{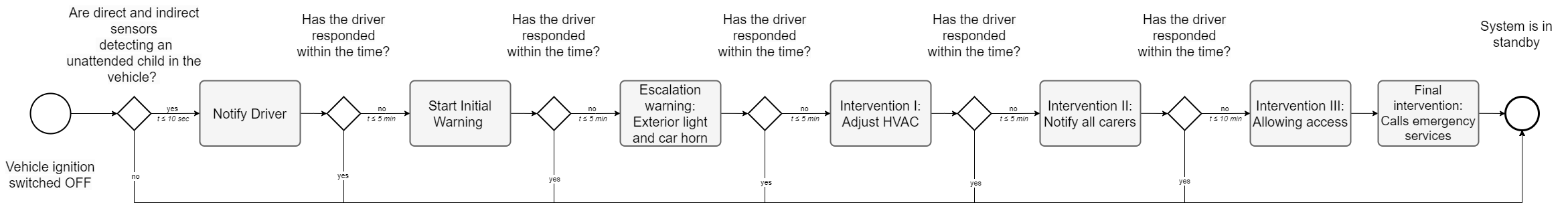}
    \caption{Escalation logic of the CPDS. The system transitions through notification and intervention stages based on time-limited driver responses.}
    \label{fig:cpd}
\end{figure}

The CPDS activation cycle initiates immediately after vehicle ignition is switched off (see Fig.~\ref{fig:cpd}). Within 10 seconds, the system evaluates sensor data to determine if a child is unattended. If detected, the driver is notified and given five minutes to respond. Without a response, CPDS escalates by activating an initial warning.

Further driver inaction triggers subsequent escalation stages, including activating exterior lights and horn signals, HVAC adjustments, notifying caregivers, unlocking doors, and ultimately contacting emergency services. Afterward, the system returns to standby, awaiting further input or intervention.

For demonstration, we focus on modeling and testing the Intervention I – HVAC adjustment step. The structured requirements for this intervention (see Listing~\ref{lst:hvac_adjustment}) serve as inputs for our GenAI-driven test case generation.

\begin{lstlisting}[style=gherkin-style,caption={HVAC Adjustment Requirements wirtten in natural language},label={lst:hvac_adjustment},
basicstyle=\scriptsize\ttfamily]
Req_CPDS_04.1. If no driver response is detected 5 minutes after escalation, the system shall adjust HVAC to maintain safe cabin conditions.
Req_CPDS_04.2. If the driver acknowledges the HVAC intervention within 5 minutes, the system shall return to standby.
\end{lstlisting}

Based on the flowchart logic (Fig.~\ref{fig:cpd}) and the requirements in Listing~\ref{lst:hvac_adjustment}, the following Gherkin test case was generated using our LLM/VLM-assisted workflow described in Section~\ref{method}.

\begin{lstlisting}[style=gherkin-style,caption={Gherkin scenario for HVAC adjustment intervention},label={lst:gherkin_hvac},
basicstyle=\scriptsize\ttfamily]
Scenario: HVAC adjustment intervention (Req_CPDS_04)
Given Vehicle.Cabin.ChildPresenceDetection.IsChildDetected is true [Req_CPDS_01.6]
And no driver acknowledgment occurs within 5 minutes of escalation [Req_CPDS_04.1]
When Vehicle.Cabin.Infotainment.HVAC.AutoOverrideActive is set to true [Req_CPDS_04.1]
And driver acknowledges within 5 minutes of HVAC override [Req_CPDS_04.2]
Then Vehicle.Cabin.ChildPresenceDetection.IsChildDetected is reset to false [Req_CPDS_04.2]
\end{lstlisting}

To maintain portability and reduce complexity, a simple base class was used to parse the Gherkin syntax without external dependencies.
Execution on the digital.auto playground required  adjustments, as some referenced VSS signals were not part of the platform’s standardized set. These were manually mapped to available standardized signals to facilitate successful execution and validation.

 \begin{lstlisting}[style=gherkin-style, caption={Python snippet for HVAC adjustment intervention test}, label={lst:python_hvac},
basicstyle=\scriptsize\ttfamily]
 class HVACAdjustmentTest(GherkinTest):
     def __init__(self, vehicle_client: Vehicle):
         super().__init__()
         self.Vehicle = vehicle_client
     async def given(self, condition):
         if condition == "Vehicle.Cabin.ChildPresenceDetection.IsChildDetected is true":
             await self.Vehicle.Cabin.ChildPresenceDetection.IsChildDetected.set(True)
         elif condition == "Vehicle.Cabin.ChildPresenceDetection.IsDriverNotified is true":
             await self.Vehicle.Cabin.ChildPresenceDetection.IsDriverNotified.set(True)
    async def when(self, action):
         if action == "Vehicle.Cabin.Infotainment.HVAC.AutoOverrideActive is set to true":
             await self.Vehicle.Cabin.Infotainment.HVAC.AutoOverrideActive.set(True)
         elif action == "Vehicle.Cabin.ChildPresenceDetection.HasDriverAcknowledged is set to true less than 5 minutes after HVAC auto override activation":
             # Simulate waiting for less than 5 minutes
             await asyncio.sleep(60)  # Simulate a short wait
             await self.Vehicle.Cabin.ChildPresenceDetection.HasDriverAcknowledged.set(True)
     async def then(self, expectation):
         if expectation == "Vehicle.Cabin.ChildPresenceDetection.IsChildDetected is reset to false":
             is_child_detected = (await self.Vehicle.Cabin.ChildPresenceDetection.IsChildDetected.get()).value
             assert is_child_detected == False, "IsChildDetected should be reset to false"
 \end{lstlisting}

\begin{figure}
    \centering
    \includegraphics[width=0.8\linewidth]{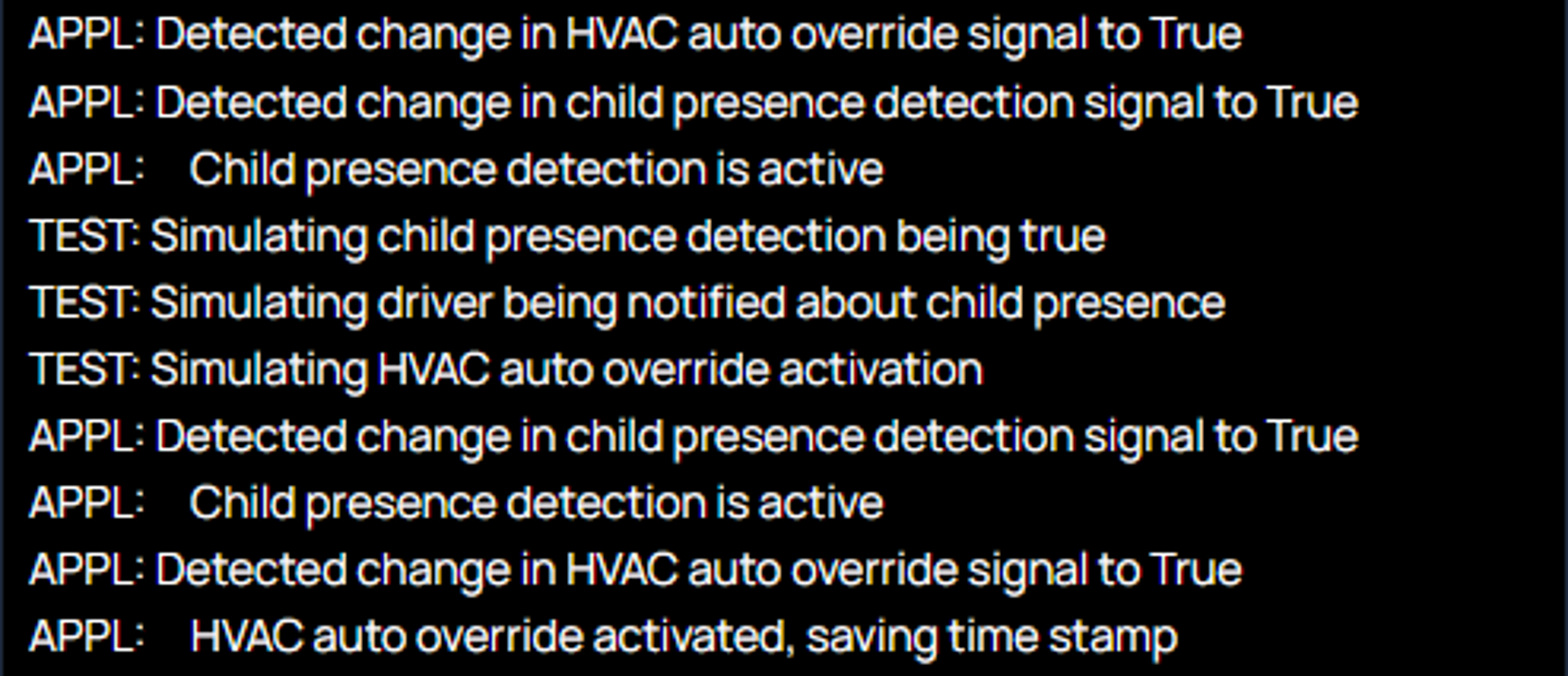}
    \caption{Console output of the executed CPDS in the digital.auto playground}
    \label{fig:cmd_output}
\end{figure}

Figure~\ref{fig:cmd_output} shows parts of the execution output, demonstrating the practical efficacy of our methodology in translating natural-language requirements and diagrams into executable test scripts suitable for SDV testing.
\section{Conclusion}
\label{conclusion}

We introduced a GenAI-driven methodology and validated its proof-of-concept implementation, demonstrating automated generation of executable test cases from automotive requirements and diagrams. Integrating LLMs, VLMs, and VSS enabled structured, reliable, and executable Gherkin and Python test scripts, significantly reducing manual specification effort.

We validated our methodology through the CPDS, a complex, safety-critical automotive use case. The evaluation demonstrated significant efficiency gains, notably reducing manual effort required for test specification and enabling rapid execution of generated tests within the digital.auto playground. However, we observed specific limitations, particularly regarding the automated generation of fully compatible scripts for the digital.auto platform, which necessitated manual adjustments and verification. Additionally, specific domain-specific signals not yet covered by standard VSS required manual mapping and validation.

Overall, our GenAI-driven approach highlights substantial potential for streamlining and automating testing processes within SDV development pipelines. Future work includes enhancing prompt engineering strategies to minimize manual intervention further, extending the VSS catalog to cover more domain-specific signals comprehensively, and overcoming compatibility constraints within the digital.auto playground. These advancements will move closer towards fully automated, reliable, and scalable test automation, significantly accelerating the development and validation cycles of next-generation automotive software.



\begin{credits}
\subsubsection{\ackname} This work has received funding from the European Chips Joint Undertaking under Framework Partnership Agreement No 101139789 (HAL4SDV) including the national funding from the Federal Ministry of Research, Technology and Space of Germany under grant number 16MEE00471K.
\end{credits}
%
%
%
\bibliographystyle{splncs04}
\bibliography{ref}

\end{document}